# Subsurface Cation Vacancy Stabilization of the Magnetite (001) Surface

**Authors:** R. Bliem,[1] E. McDermott,[2] P. Ferstl,[3] M. Setvin,[1] O. Gamba,[1] J. Pavelec,[1] M. A. Schneider,[3] M. Schmid,[1] U. Diebold,[1] P. Blaha,[2] L. Hammer,[3] G.S. Parkinson[1]*

**Affiliations:**

[1]Institute of Applied Physics, Wiedner Hauptstrasse 8-10, Vienna University of Technology, 1040 Vienna, Austria.

[2] Institute of Materials Chemistry, Getreidemarkt 9, Vienna University of Technology, 1060 Vienna, Austria.

[3]Chair of Solid State Physics, University of Erlangen-Nürnberg, Staudtstrasse 7, 91058 Erlangen, Germany.

*Correspondence to: parkinson@iap.tuwien.ac.at

**Abstract**: Iron oxides play an increasingly prominent role in heterogeneous catalysis, hydrogen production, spintronics and drug delivery. The surface or material interface can be performance limiting in these applications, so it is vital to determine accurate atomic-scale structures for iron oxides and understand why they form. Using a combination of quantitative low-energy electron diffraction, scanning tunneling microscopy, and density functional theory calculations, we show that an ordered array of subsurface iron vacancies and interstitials underlies the well-known ($\sqrt{2}\times\sqrt{2}$)R45° reconstruction of $Fe_3O_4(001)$. This hitherto unobserved stabilization mechanism occurs because the iron oxides prefer to redistribute cations in the lattice in response to oxidizing or reducing environments. Many other metal oxides also achieve stoichiometric variation in this way, so such surface structures are likely commonplace.

**Main Text:** The properties of metal oxide surfaces are inextricably linked to their atomic scale structure, which makes obtaining a precise structural model a vital prerequisite to understanding and modeling surface processes. The typical starting point to guess the structure of a metal oxide surface is the Tasker surface polarity criteria (*1*). These simple and convenient rules have proven remarkably successful for predicting the stable terminations of model systems such as $TiO_2$, MgO, and ZnO (*2, 3*). A second guiding principle to emerge from 20 years of metal oxide surface science is the vital role played by oxygen vacancies ($V_{OS}$). The surface $V_O$ concentration is usually in equilibrium with the bulk, where such defects form when the oxide is reduced. At the surface, $V_{OS}$ strongly affect the electronic structure, are active sites for chemical reactions (*4, 5*), and play a central role in surface reconstructions. There exists another class of metal oxides however, where stoichiometric variation is accommodated primarily via the cations; the iron oxides are the prototypes. Wüstite (FeO), magnetite ($Fe_3O_4$), and hematite ($\alpha$-$Fe_2O_3$) are all based on a close packed $O^{2-}$ anion lattice, and differ primarily in the number, oxidation state, and distribution of the cations among octahedral and tetrahedral interstitial sites, $Fe_{oct}$ and $Fe_{tet}$ (*6*). Near the phase boundaries, FeO can incorporate up to 17% of $Fe_{oct}$ vacancies, while the metastable phase maghemite ($\gamma$-$Fe_2O_3$) is essentially $Fe_3O_4$ with 1/6 of the $Fe_{oct}$ removed. We now show that an ordered array of subsurface cation vacancies and interstitials underlies the well-known ($\sqrt{2}\times\sqrt{2}$)R45° reconstruction of $Fe_3O_4(001)$. This stabilization is driven by the material adopting a stoichiometry compatible with its environment. The oxides of Co, Mn, and Ni exhibit similar cation redistributions in reducing and oxidizing environments, so such reconstructions may be a common occurrence.

In the (001) direction, $Fe_3O_4$ consists of alternating planes of $Fe_{tet}$ and $Fe_{oct}$-O atoms and thus appears a classic Tasker type 3 polar surface (*1, 3*). Indeed, early studies of the $(\sqrt{2}\times\sqrt{2})R45°$ reconstruction considered only structures compatible with polarity compensation (*7-9*). However, $Fe_3O_4$ is metallic, so it is not clear why the polar catastrophe should occur at all. In fact, the current model for the $Fe_3O_4$(001) surface is based on a nominally polar bulk truncation at the $Fe_{oct}$-O plane. For this termination (Fig. 1A), density functional theory (DFT) (*10-12*) finds pairs of surface $Fe_{oct}$ relaxed in opposite directions perpendicular to the row. This structure produces the requisite $(\sqrt{2}\times\sqrt{2})R45°$ periodicity and is in qualitative agreement with undulating rows of $Fe_{oct}$ atoms observed in scanning tunneling microscopy (STM) images (*7, 8*). The emergence of orbital order among the $Fe_{oct}$ cations in subsurface layers has been likened to the "Verwey" phase of $Fe_3O_4$ (*11*), which forms below a transition temperature of 125 K. This structure has been widely accepted, largely on the basis of low-energy electron diffraction (LEED *IV*) (*12*) and surface x-ray diffraction (SXRD) (*10*) measurements, which purport to confirm the structure. However, the quantitative measure of agreement between experimental and theoretical LEED *IV* curves, the Pendry *R*-factor ($R_P$) (*13*), was poor ($R_P = 0.34$) in comparison to what is achieved for metal and elemental semiconductor surfaces ($R_P < 0.2$). This classic study, and several others (*14-16*), suggest that the achievable $R_P$ is limited for metal oxide surfaces, either because the scattering calculations fail to account for ionic bonding (*14-16*), or because oxides contain many defects (*12*).

Perspective (Fig. 1B) and top (Fig. 1C) views of the subsurface cation vacancy (SCV) structure of $Fe_3O_4$(001), as determined by DFT+U calculations, show that the $(\sqrt{2}\times\sqrt{2})R45°$ periodicity arises from replacing two $Fe_{oct}$ from the third layer by an interstitial $Fe_{int}$ with tetrahedral coordination in the second layer; a net removal of one cation per $(\sqrt{2}\times\sqrt{2})R45°$ unit cell. The subsurface reorganization distorts the surface layer, and the resulting undulations in the $Fe_{oct}$ rows are more pronounced than in the bulk truncation model (compare Figs. 1A and 1B). All cations in the four outermost layers possess a magnetic moment of 3.98-4.01 $\mu B$. This is indicative of an $Fe^{3+}$-like character that is also observed in x-ray photoelectron spectroscopy (XPS) (*9*); for angle-resolved XPS see Fig. S1 (*17*). The $Fe_{oct}$ and $Fe_{tet}$ sublattices remain antiferromagnetically coupled as in the bulk, yielding a reduced net magnetic moment in the SCV reconstruction. The absence of $Fe^{2+}$ suggests that the contribution of the orbital moment should be small. Table S1 (*17*) contains the magnetic moments of all atoms in the surface region together with their optimized DFT+U coordinates. The energetic preference for the SCV structure is borne out by atomistic thermodynamics calculations (Fig. 1E) (*18*). The distorted bulk truncation is only favored below an oxygen chemical potential of -3 eV, corresponding to an $O_2$ partial pressure of $<10^{-20}$ mbar at 900 K. Under such reducing conditions, however, Fe-rich surface phases occur (*19*). While $\alpha$-$Fe_2O_3$ is predicted to take over as the thermodynamically stable phase above $\approx 10^{-9}$ mbar, the $(\sqrt{2}\times\sqrt{2})R45°$ reconstruction persists up to $10^{-5}$ mbar. It has been shown that $\alpha$-$Fe_2O_3$ inclusions grow slowly at $10^{-6}$ mbar $O_2$ (*20*), while conversion of the surface to $\gamma$-$Fe_2O_3$(001) requires extremely oxidizing conditions (*9*).

A key success of the SCV structure is that it explains the extraordinary thermal stability of Au, Ag and Pd adatoms at the $Fe_3O_4$(001), which resist agglomeration into clusters up to 700 K (*19, 21-23*). Adatoms exclusively occupy just one of the two available bulk continuation sites per $(\sqrt{2}\times\sqrt{2})R45°$ unit cell. This finding is difficult to reconcile with the distorted bulk truncation model, because the two sites are identical save for the subtle relaxations and the subsurface orbital ordering. Unsurprisingly, DFT+U calculations for Au adatom adsorption on the bulk-truncated surface reveal no clear preference for one specific site. On the SCV structure the interstitial $Fe_{int}$ blocks the adsorption of an adatom in the bulk continuation site directly above. For Au, a stable binding configuration (2.03 eV) is only found in the two-fold coordinated site located above the "bulk-like" subsurface region (see Fig. 1D), which results in nearest neighbors 8.4 Å apart (*19, 21-23*).

To independently confirm the SCV structure of $Fe_3O_4(001)$ we conducted LEED *IV* experiments. Figure 2A shows selected experimental spectra, together with best fit curves resulting from optimized structural parameters. (See the Supplement (*17*) for the full set of curves (Fig. S3), the optimized coordinates (Table S2), their associated error estimates, and a structure file that can be opened by most visualization software programs.) The best-fit structure achieved a Pendry R-factor of $R_P = 0.125$, a vast improvement over the 0.34 obtained previously for the distorted bulk truncation model (*12*). Moreover, the agreement between the LEED *IV* parameters and the DFT+U optimized structure is excellent (see Table S3 (*17*)). In the surface layer, all atoms relax toward the nearest interstitial $Fe_{int}$ atom in the second layer (see Fig. 1, B and C). The largest relaxation (0.28 Å in the surface plane) occurs for the O furthest from $Fe_{int}$. In DFT+U calculations this atom has a large magnetic moment (0.35 $\mu_B$), and is close to a -1 charge state. The lateral relaxations of the surface $Fe_{oct}$ atoms (0.11 Å) produce a peak-to-peak undulation of 0.22 Å in the $Fe_{oct}$ rows. While the $Fe_{int}$ atom resides 0.08 Å below the bulk-like $Fe_{tet}$ atoms, a slight rumpling in the neighboring $Fe_{oct}$-O layers preserves the Fe-O bond lengths within 2 % (1.889 Å).

Because STM is the method of choice for monitoring dispersed adsorbates, it is important to establish the appearance of adsorption sites in the SCV structure. Three empty-states STM images taken at the same area with different tunneling voltages are shown in Fig.3, A to C. An Fe adatom (a typical defect following sputter/anneal cycles (*24*)) in the upper left corner served as a marker. Fe occupies the same adsorption site as Au, Ag and Pd adatoms (*19, 21-23*), i.e., the bulk continuation site marked with the yellow × in Fig. 1C. At low sample bias, the undulations in the $Fe_{oct}$ row appeared pronounced and the × is located where the Fe rows come closest together. [In our previous work (*19, 21-23*) we called this position the 'narrow site'.] When the bias was increased to 2.0 V, the $Fe_{oct}$ rows appeared almost straight, and it was difficult to discern the $(\sqrt{2}\times\sqrt{2})R45°$ periodicity. At 3.4 V, the undulations returned, but the × was located in the opposite phase of the reconstruction, i.e., where the $Fe_{oct}$ rows were furthest apart.

This trend was borne out in STM simulations (Fig. 3, D to F), and can be understood by examining the density of states (DOS) of the surface $Fe_{oct}$ atoms (Fig. 3J) and the corresponding charge density contour plots (Fig. 3, G to I). At positive sample bias, electrons tunnel from the tip into empty states above the Fermi level. The *d* states are split by the crystal field, and the $t_{2g}$ orbitals straddle $E_F$. At 0.17 V, electrons can tunnel only into a $t_{2g}$ orbital, which is tilted off-axis because of an antibonding interaction with the underlying O-atom (Fig. 3G). Thus the imaged electron density is not symmetric about the atom cores. At 3.4 V, when tunneling is dominated by the symmetric $d_{z^2}$ and $d_{x^2-y^2}$ $e_g$ orbitals, the STM images are more in line with the true relaxations as determined by DFT+U and LEED *IV*.

The SCV structure described here could not have been predicted using polarity compensation rules (*3*). The driving force of the reconstruction is the oxygen chemical potential, consistent with previous reports of reduced terminations on the $Fe_3O_4(111)$ (*25, 26*) and $Fe_2O_3(0001)$ (*27*) surfaces. Although the bulk is restricted to the FeO (rocksalt), $Fe_3O_4$ (spinel) and $Fe_2O_3$ (corundum) structures, the lack of 3D periodicity at the surface permits the formation of the SCV structure, a distinct phase with an intermediate $Fe_{11}O_{16}$ stoichiometry. In this light, the $(\sqrt{2}\times\sqrt{2})R45°$ reconstruction must be seen as the first stage in the oxidation of $Fe_3O_4$. With all cations in a bulk-like environment it is a highly stable configuration. It is unlikely that the structure penetrates deeper, as the interstitial $Fe_{int}$ prevents the occupation of a bulk $Fe_{tet}$ site in the next layer. The same effect is at work in the surface layer (Fig. 1D), where the $Fe_{int}$ blocks Au adsorption in one of the bulk continuation $Fe_{tet}$ sites.

The $R_P$ of 0.125 achieved in our LEED *IV* experiments strongly supports the SCV structure, and also shows that there is no inherent problem in the LEED *IV* methodology that limits the obtainable agreement for complex metal oxide surfaces. While it is of course important to ensure the surface under investigation is of the highest quality (ideally cross-checked by in-situ imaging), and to minimize damage by the LEED electron beam, using the correct trial structure in the calculations is essential. To reinforce this point, we also analyzed the experimental dataset acquired in a previous LEED *IV* study of this surface (*12*). Despite the crystals, vacuum system, measurement temperature, and preparation conditions differing between the measurements, a similar $R_P$ of 0.124 was obtained.

Understanding the true structure of the $Fe_3O_4$(001)-($\sqrt{2}\times\sqrt{2}$)R45° surface is vital to correctly interpret experimental results acquired for this surface. For example, the observation of a surface band gap was taken as evidence for a "surface Verwey transition" (*11*). Our results show that the surface band gap is in fact due to an entirely different iron oxide phase at the surface, and is not related to charge/orbital ordering. The SCV reconstruction also likely explains why spin polarized photoemission does not yield the 100 % spin polarization predicted for the bulk (*28*). This surface-sensitive technique mainly probes the $Fe_{11}O_{16}$ phase, which has little DOS at $E_F$, no pronounced spin polarization (see Fig. 3J), and a reduced net magnetization. Such a "magnetic dead layer" has been blamed for the poor performance of $Fe_3O_4$ based spintronics devices (*29, 30*). Perhaps the most exciting consequence of the SCV structure is that it supports isolated metal adatoms that remain stable against agglomeration until the reconstruction is lifted at 700 K (*23, 31*). This distinctive property, which represents the ultimate limit of the adsorption template phenomenon, provides a well-defined initial state for fundamental studies of cluster nucleation and growth (*22, 23*), and offers an ideal model system to uncover the mechanisms of single atom catalysis (*32*).

Finally, given that the oxides of Co, Mn, and Ni exhibit a similar tendency to rearrange the cation lattice as the iron oxides, and even traverse the same oxidation pathway (from rocksalt structures through spinel to corundum), cation vacancy reconstructions are likely commonplace.

**Acknowledgments:** GSP and OG acknowledge support from the Austrian Science Fund project number P24925-N20. RB and EM acknowledge a stipend from the TU Vienna and Austrian Science Fund doctoral college SOLIDS4FUN, project number W1243. UD acknowledges support by the ERC grant "OxideSurfaces". PB was supported by the Austrian Science Fund project SFB-F41 ViCoM. All authors acknowledge Prof. Z. Mao and T.J. Liu (Tulane University) for the synthetic sample used in this work. We thank Prof. W. Moritz for sharing experimental LEED *IV* spectra.

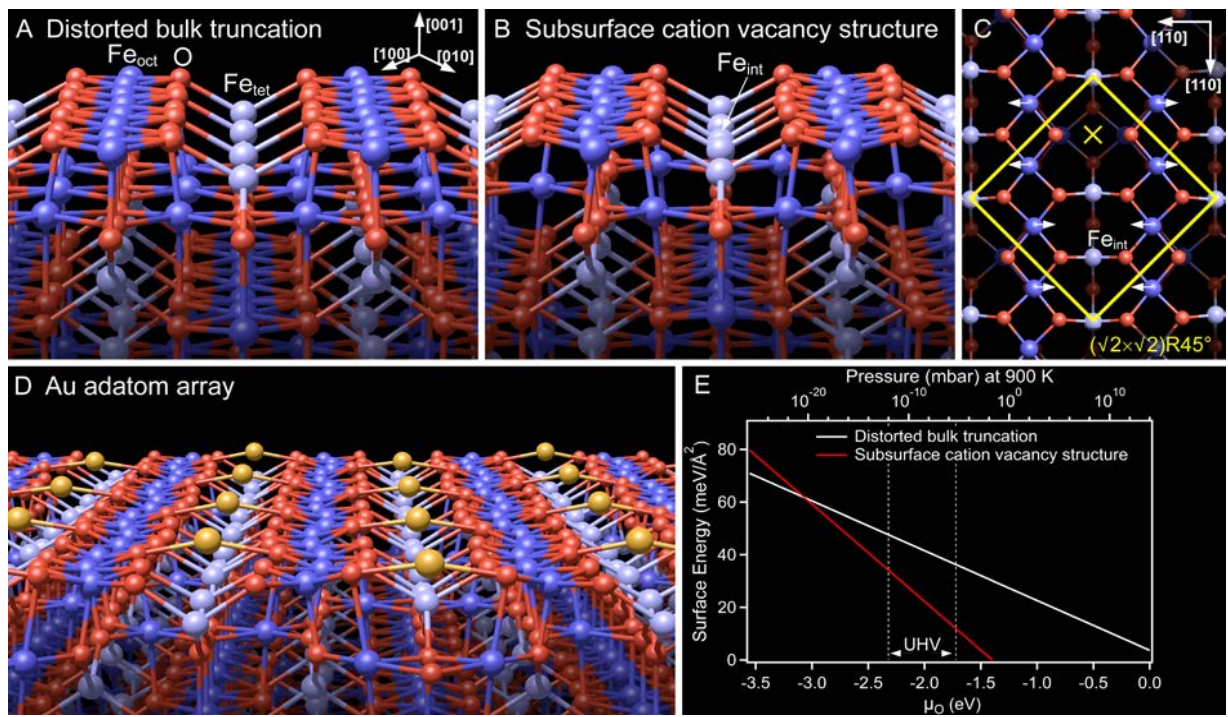

**Fig. 1**. Density functional theory based calculations show that an ordered array of cation vacancies underlies the (√2×√2)R45° reconstruction of the $Fe_3O_4$(001) surface. (A) Minimum energy structure of the distorted bulk truncation model (*10, 12*). (B and C) Perspective and plan views of the subsurface cation vacancy structure. A pair of $Fe_{oct}$ cations from the third layer is replaced by an interstitial $Fe_{tet}$ (labeled $Fe_{int}$) in the second layer. (D) Au adatoms bind strongly (2.03 eV) to surface O at positions without second-layer $Fe_{tet}$ atoms. These sites, are marked with an × in panel C, result in a nearest-neighbor Au distance of 8.4 Å (*21*). (E) Surface energy as a function of oxygen chemical potential, $\mu_O$, for both structures. The $\mu_O$ is converted to an $O_2$ pressure for a nominal annealing temperature of 900 K. The cation vacancy structure is stable in all experimentally accessible conditions.

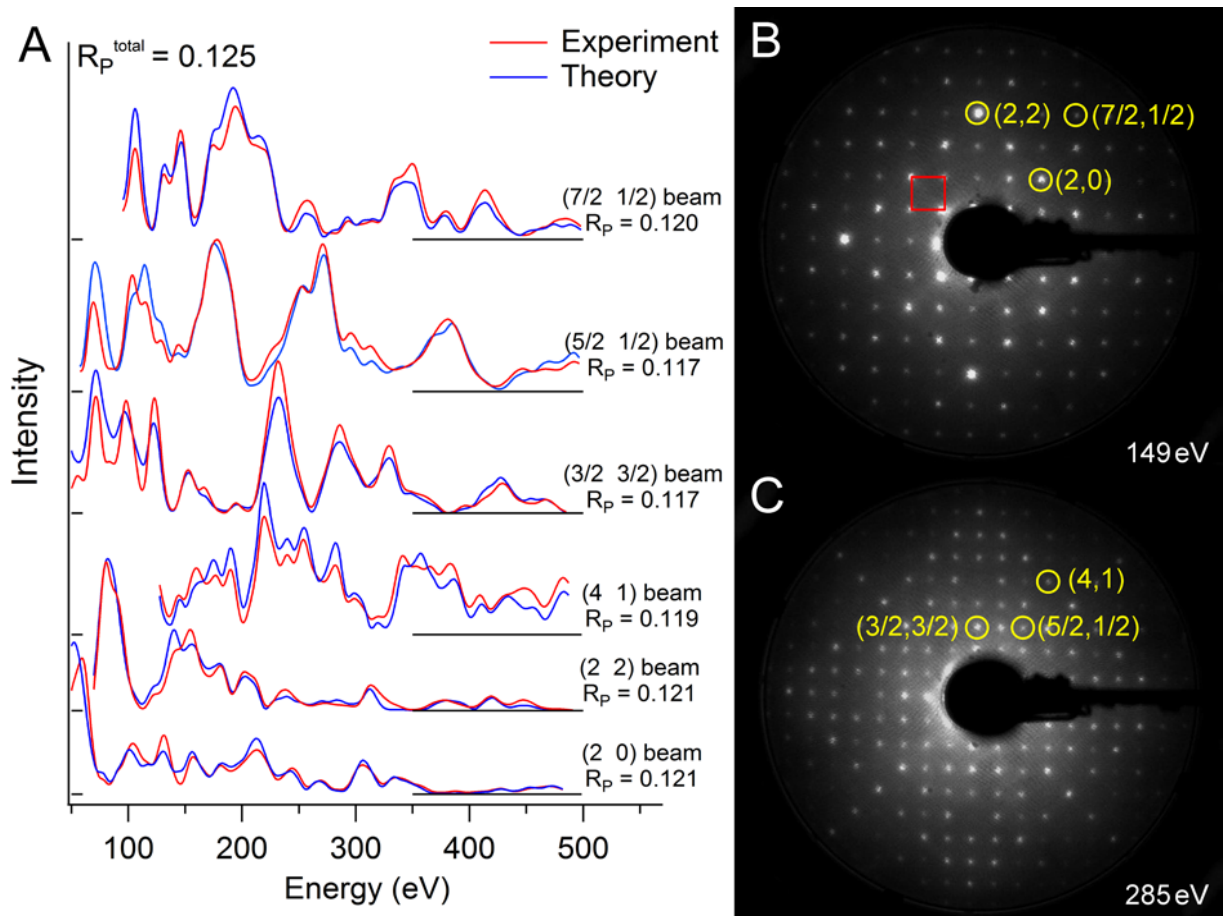

**Fig. 2.** Quantitative low energy electron diffraction measurements unambiguously confirm the subsurface cation vacancy termination of $Fe_3O_4$(001). (A) Comparison of selected experimental LEED *IV* spectra and theoretical curves for the optimized subsurface vacancy structure. The final $R_P$ for the best fit-structure is 0.125; the selected beams have an individual $R_P$ close to this average. (B and C) Experimental LEED patterns for electron energies of 149 eV and 285 eV. Diffraction spots highlighted with yellow circles correspond to the curves in (A). The red square indicates the $(\sqrt{2}\times\sqrt{2})R45°$ unit cell.

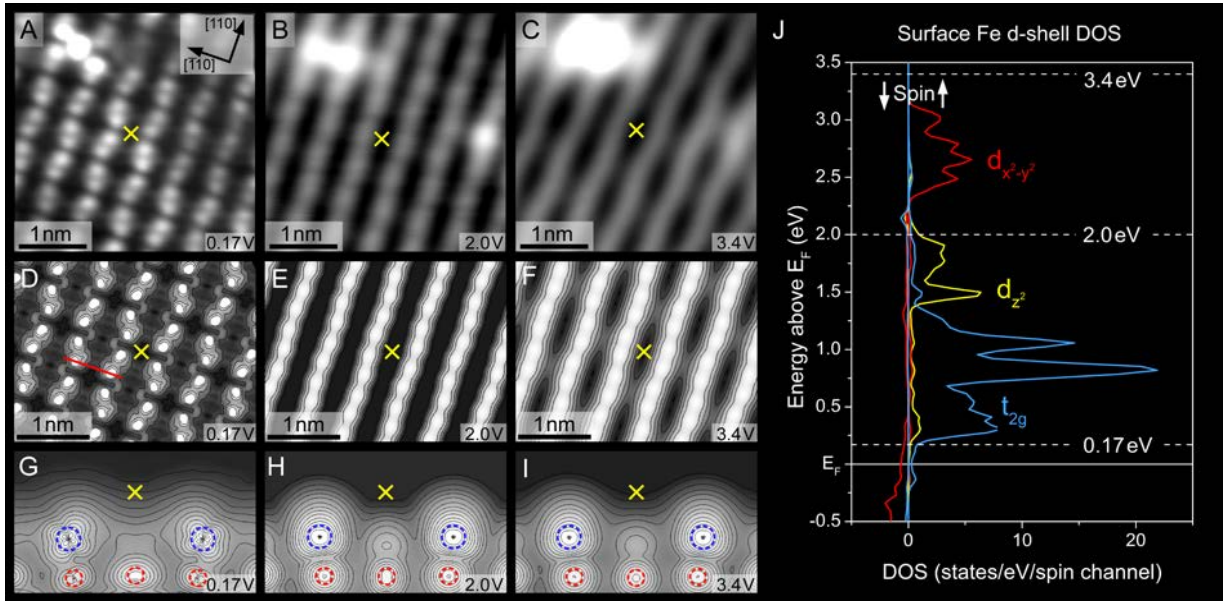

**Fig. 3.** STM images of the Fe$_3$O$_4$(001) surface are dominated by electronic effects. (A to C) Empty states STM images obtained on the same sample area in constant-height (A) and constant-current mode with $I_t$ = 0.15 nA (B) and $I_t$ = 0.15 nA (C). The bias voltages are given in each panel. An Fe adatom defect in the top left of each image defines a constant position (×). As the STM bias is increased the apparent undulations in the Fe$_{oct}$ rows reverse phase. (D to F) Simulated STM images at scanning conditions corresponding to A to C. (G to I) Density-of-States contour plots in cross section at the position indicated by the red line in (D). The t$_{2g}$ orbitals of the surface Fe$_{oct}$ atoms (blue dashed circles) are distorted resulting in a density that is asymmetric about the atom core position. (J) Electronic density of states above E$_{Fermi}$ for the surface Fe$_{oct}$ atoms. As the tunneling bias is increased, different Fe *d* orbitals contribute to the STM images. At 0.17 V, only part of the t$_{2g}$ orbital is accessed, followed by more spherically symmetric d$_{z^2}$ and d$_{x^2-y^2}$ e$_g$ orbitals at higher bias. The higher bias gives a more realistic indication of the atomic core positions.

**Supplementary Materials:**

Materials and Methods

Figs. S1 to S3

Tables S1 to S3

Captions for database S1

References (*33-35*)

**Supplementary Materials:**

**Materials and Methods:** The STM experiments were performed at 78 K on a synthetic $Fe_3O_4$ single crystal prepared by cycles of sputtering with $Ar^+$ ions ($I_{sample}$ = 1 µA, 1 keV, 15 min) followed by annealing at 930 K in $10^{-6}$ mbar $O_2$ for 30 min. The STM setup is a two-chamber UHV system with base pressures below $10^{-11}$ mbar, equipped with an Omicron LT STM. In the LEED setup, the sample was annealed at 920 K in an $O_2$ background of $10^{-8}$ mbar using an $O_2$ doser that enhanced the pressure at the sample by a factor ≈500. XPS measurements were obtained at room temperature using a SPECS FOCUS 500 monochromatic x-ray source (Al K$\alpha$ x-rays) and a SPECS PHOIBOS 150 electron analyzer with a pass energy of 15 eV.

The LEED experiments were performed using a conventional LEED optics. Spot intensities were recorded in the energy range 50 - 500 eV for normal incidence (within ~0.1°) of the primary beam. The data set was acquired at 90 K, as temperature causes no significant structure-related effects in LEED (*12*). All accessible symmetrically equivalent beams were measured, averaged and moderately smoothed in order to remove any residual experimental noise. The total database consists of 40 symmetrically inequivalent beams (23 integer and 17 fractional order) with an accumulated energy range of 11308 eV.

The intensity calculations were performed using the TensErLEED program package (*33*). Atomic scattering was described by 14 phase shifts calculated for neutral scatterers (*34*). Electron attenuation was simulated by an optical potential ($V_{0i}$ = 5.75 eV). The energy dependent real part of the inner potential takes the form $V_{0r}$ = max[-9.94, -0.07-75.63/$\sqrt{(E+20.16)}$] eV, the constant part was adjusted within the fit to be $V_{00}$ =-2.0 eV.

For the LEED analysis $Fe_3O_4$ was assumed to be in its cubic spinel structure disregarding the small distortions imposed by the Verwey transition. All symmetry-allowed geometrical parameters within the outermost four layers were varied in the final fit, as well as all vertical parameters down to the sixth layer. The vibrational amplitudes of the first $Fe_{tet}$ and $Fe_{oct}$-O layer atoms were also optimized. This resulted in a total of 59 geometrical and 5 vibrational parameters under variation. The large data base (11308 eV) results in a redundancy factor of 8.3, so that any overfitting of the data could be excluded. The Pendry R-factor (*13*) was used for a quantitative comparison of experimental and calculated spectra. The best fit achieved a Pendry *R*-factor $R_{min}$=0.125. The error bars of the best fit configuration were determined via the variance of the R-factor value var($R_{min}$) = 0.008 (*13*).

The DFT calculations were performed using the full-potential augmented plane wave + local orbital (APW+lo) method as implemented in WIEN2k (*35*). We use the generalized gradient approximation with a Hubbard U ($U_{eff}$ = 3.8 eV) to treat the highly correlated Fe *3d* electrons. The surface phase diagram was constructed using the formalism described by Reuter *et al.* (*17*). Magnetite surface reconstructions were modeled on a 17-layer slab with inversion symmetry and a vacuum layer of 25 bohr, built from a ($\sqrt{2}\times\sqrt{2}$) supercell of bulk cubic magnetite. The total energy of bulk magnetite and Fe was calculated with lattice parameters optimized for $U_{eff}$ = 3.8 eV. Molecular oxygen was optimized in a cube with edge length of 34 bohr. In each model, atomic sphere sizes of 1.86, 1.15, and 2.1 bohr were used for Fe, O, and Au, respectively. A plane-wave cutoff of 23 Ry was used, and the 2D Brillouin zones of the surface models were sampled with a 3×3×1 k-mesh. The surface models were relaxed until all forces were below 1 mRy/bohr. A Fermi broadening of 0.006 Ry was employed.